\documentclass[a4paper]{article}

\usepackage{graphicx}

\begin{document}

\title{PARTICLE DARK MATTER: \\THE STATE OF THE ART}

\author{Marco Regis\\{\it Astrophysics, Cosmology \& Gravitation Centre, } \\
{\it Department of Mathematics \& Applied Mathematics,} \\{\it University of Cape Town,}\\
 {\it Rondebosch 7701, Cape Town, South Africa}\\  and \\ {\it Centre for High Performance Computing,} \\
{\it 15 Lower Hope St, Rosebank, Cape Town, South Africa}}

\date{}

\maketitle

\begin{abstract}

Although well established observations on cosmological, cluster, and galactic scales strongly suggest the existence of dark matter (DM), our understanding of its non-gravitational properties is still lacking.
I review basic aspects of particle dark matter and detection strategies, outlining the state of the art for searches in direct detection experiments, indirect observations, and particle production at colliders.  
A particular focus is dedicated to recent experimental results which could have provided hints for unveiling the DM nature.

\end{abstract}

\section{Introduction}
One of the gold-rushes in cosmology and particle physics involves unveiling the elusive nature of dark matter (DM).
Predicting when and through which method this hunt will be successful is not a simple task.
Somehow, the discovery of DM can be compared to the eruption of a volcano.
To predict the latter is quite difficult as well, and geologists typically rely on the observations of anomalous gas emissions and small earthquakes in the volcano area. So, perhaps, to know if the discovery of DM is actually close in time, the right question could be: have we seen anomalous excess in some experiments and consequent small earthquakes in the particle physics community?
Recent results from DAMA, PAMELA, CDMS, and COGENT certainly share such symptoms.
In the following, after a general introduction to particle DM and related detection strategies, I will critically discuss what they can tell us about the DM nature.

In the last three decades, gravitational evidences for dark matter have been accumulated at the galactic, cluster and cosmological scales.
Historically, the first solid hint for its existence came from rotation curves in spiral galaxies, e.g., (Bosma, 1981), (although first claims date back to 30's (Zwicky, 1933) and refer to cluster dynamics). On galactic scales, the evidence for DM is quite compelling and mostly relies on estimates of kinematic mass from observations of rotation velocity of stars in rotationally supported galaxies and velocity dispersion in pressure supported galaxies (and assuming Newtonian gravity).
On cosmological scales, our understanding have recently experienced tremendous progresses, allowing to distinguish among many different cosmological models.
As a cornerstone, the measurement of the power spectrum of cosmic microwave background (CMB) anisotropies led to a detailed determination of cosmological parameters, e.g., (Komatsu et al, 2010). It is in agreement with large scale structure (LSS) and Big Bang nucleosynthesis (BBN) data, and they all require the matter density in the universe to be much larger than the baryonic density.
To infer the mass of a cluster one can rely on different estimators, such as the dispersion velocity of galaxies in the cluster, gravitational lensing and thermal X-ray emissions (which provides the temperature of the hot intra-cluster gas and so the hydrostatic pressure).
They are in good agreement, leading to $\Omega_m \sim 0.2$, which is consistent with cosmological constraints.

These gravitational evidences do not fully shed light on the microscopic properties of DM. On the other hand, the consistency of this scenario points toward collisionless and dissipationless DM, as in particular required by formation of galactic halos and by clusters merger dynamics (e.g., the so-called Bullet cluster (Clowe et al., 2006)).


\section{Particle Dark Matter}
{\bf Baryonic}

Since a dissipative form of matter would condense without forming extended halos in galaxies, the most plausible baryonic DM is in the form of massive astrophysical compact halo object (MACHO), rather than elementary particles.
They are macroscopic objects which do not produce a significant amount of observable radiation, and typically form before BBN, in order not to affect the light element abundances (e.g., primordial black holes).
Gravitational microlensing can be used to detect MACHOs and the rate of gravitational microlensing of stars in the Small and Large Magellanic
Clouds constrains the mass fraction of MACHOs in the Milky Way halo to be $< 20\%$, in case of masses between $6\cdot 10^{-8} - 15\, M_\odot$ (Tisserand et al., 2006).
Other analysis share the same common results, namely, that, assuming Newtonian gravity, a significant amount of non-baryonic DM seems to be unavoidable.

{\bf Non-Baryonic}

Depending on the velocity of the particle at the time when galaxy structures could start to form, the DM has been termed hot, warm, and cold.
A collisionless relativistic (i.e., hot) species tends to erase fluctuation below its free-streaming length, and leads to a top-down hierarchy in structure formation,
with galaxies and clusters formed through a process of fragmentation.
This possibility is currently strongly constrained, and hot dark matter (HDM) has been ruled out as the main DM component, $\Omega_{HDM}h^2\leq0.01$ (Komatsu et al., 2010).
The cold dark matter (CDM) paradigm plus cosmic inflation form the basis for the standard cosmology.
Although the CDM scenario is very successful on large scale, and the bottom-up approach can explain cluster formation and distribution of stars in galaxies, potential disagreements with the naive expectations of the theory may be present at small scales.
In order to alleviate them, warm dark matter (WDM) has been suggested. The term "warm" labels DM candidates with velocity dispersion and free streaming length standing in between CDM and HDM. For this reason, fluctuations on small scales are suppressed, reducing the formation of small structures.
WDM is a fully viable hypothesis, but on the other hand, the DM mass window allowed by cosmological constraints (including Ly$\alpha$ observations, phase-space distribution function arguments, and radiative X-ray emissions) has been shrinking and shrinking.

For all the reasons outlined above, we will focus on dissipationless collsionless non-baryonic cold dark matter.

Making the conservative assumptions that the DM is stable and was in thermal equilibrium with the plasma of the primordial universe, its relic density can be expressed through (e.g., (Kolb \& Turner, 1990)):
\begin{equation}
\Omega_{DM}h^2\simeq \frac{3\cdot10^{-27}{\rm cm^3s^{-1}}}{<\sigma_a\,v>}\;,
\label{eq:miracle}
\end{equation}
where $<\sigma_a\,v>$ is the thermally averaged annihilation rate. Eq.~\ref{eq:miracle} implies that DM particles with annihilation cross section mediated by weak interactions (and mass $m_{DM}\sim\mathcal{O}$(100 GeV)) are naturally produced with the correct thermal relic density. This is the so called WIMP miracle and the appeal of the weakly interacting massive particle (WIMP) class of DM candidates mostly stems from it.
Theories beyond the standard model (SM) of particle physics can easily account for WIMPs. Indeed, one of their foundation motivations relies on addressing electro-weak (EW) symmetry breaking issues of the SM, so they naturally introduce particles at EW scale and with weak couplings. Examples include lightest supersymmetric particles in Supersymmetry (Jungman et al., 1996), and Kaluza-Klein states in flat and warped extra-dimension models (e.g., (Regis, Serone \& Ullio, 2007) and (Panico et al., 2008)).
Moreover, weak couplings ensure that its interaction with standard matter is sizable which implies that prospects for detection in current and near future experiments are very promising. 

From a dimensional analysis $\Omega_{DM}h^2\propto <\sigma_a\,v>^{-1}\sim m_{DM}^2/g_{DM}^4$, where $g_{DM}$ is the coupling constant.
Recent experimental results (that I will present in the next Section) may require a mass significantly larger or smaller than the canonical $m_{DM}=100$ GeV (which can be obtained considering $g_{DM}=g_w\simeq0.65$), which then leads to $g_{DM}\neq g_w$.
This is the case of the so called WIMPless scenario (Feng, 2010) which can share most of the nice properties of WIMPs, although requiring the introduction of an ad hoc gauge group providing the coupling $g_{DM}$.

In the rest of the talk, I will keep the discussion general, describing observational prospects for DM candidates with mass in the GeV-TeV regime and cross-sections of order of magnitude not too far from the weak case.

\section{Detection Strategies}
\subsection{Production at Colliders}
The production of WIMPs at collider stems from the process:$$p_{SM}+p_{SM}\rightarrow \chi_{DM}+\chi_{DM}+"some\,p_{SM}"\;,$$ where $p_{SM}$ is some well-known particles of the SM.
For $m_{DM}\leq$ few TeV, DM particles $\chi_{DM}$ will be produced at LHC and ILC.
On the other hand, the weak interaction makes them invisible and one should disentangle the signal in $"some\,p_{SM}"$ from the background.
It can be quite challenging at LHC where, most probably, WIMPs can be only detected as missing energy events.
Since this signature is common to most exotic physics beyond the SM, it has to be combined with other observables (e.g., DM cosmological relic density, or direct and indirect searches results) to lead to a robust DM discovery.
Tests of the particle physics framework in which the DM candidate is embedded can provide other indirect cross-checks for DM interpretations.
Currently, LHC is in a commissioning phase, starting to see first physics results, which however involve "rediscovery" of SM, while a new discovery of any sort is rather unlikely in the forthcoming months (see (Straessner, 2010)).

\subsection{Direct Detection}
As mentioned before, the DM is not only a cosmological issue, but rather it is postulated down to galactic scales and a significant WIMP population is expected at our location in the Milky Way.
Because of WIMP miracle, and using crossing symmetry, scatterings of WIMPs with ordinary matter proceed through weak interactions and
the direct detection strategy consists in recording the recoil energy of target atomic nuclei after scattered by a WIMP.
To have an handle on the dependencies of the signals on the dark matter and target properties one can consider the recoil energy of the nucleus in the laboratory frame (given by non-relativistic kinematics), $E_r=|{\bf q}|^2/(2\,m_N)=\mu^2 v^2(1-\cos \theta)/m_N$  and a rough approximation for the events rate $R\sim N\,\sigma_a v\,\rho_{DM}/m_{DM}$ where explicitly the WIMP mass $m_{DM}$, local density $\rho_{DM}$, scattering cross section $\sigma_s$, and velocity $v$ relative to the target, and number $N$ and mass $m_N$ of target nuclei appear (with $\mu \equiv m_N m_{DM}/(m_N+m_{DM}) $ being the reduced mass and $\theta$ the scattering angle in the center-of-mass system). 
The nuclear recoil produced by the WIMP scattering can be measured by detecting the induced light, charge or phonons through scintillation, ionization and lattice heat. Current direct detection experiments exploit one or combine two of such techniques.

\begin{figure}  
 \begin{minipage}[htb]{7cm}
   \centering
   \includegraphics[width=7cm]{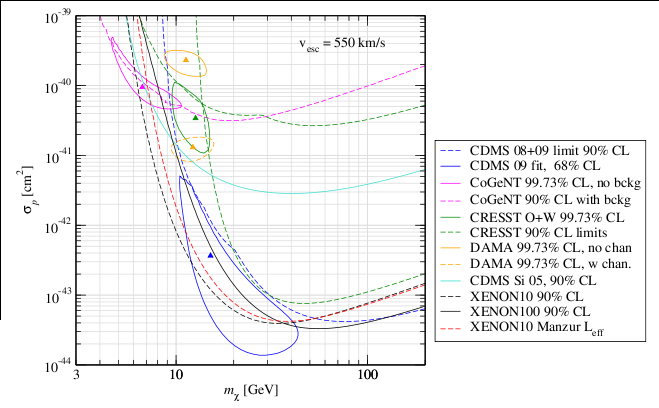}
 \end{minipage}
\begin{minipage}[htb]{5cm}
\includegraphics[width=5cm]{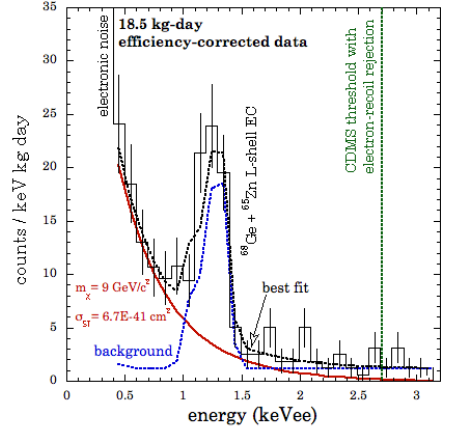}
 \end{minipage}
\caption{{\it Left:} WIMP parameter space selected by direct detection experiments in the case of spin-independent elastic scattering. From (Schwetz, 2010).
{\it Right:} COGENT results (Aalseth et al., 2010) (counts versus recoil energy). }
\label{fig:direct}
\end{figure}

In the following I outline recent results and their possible interpretations in terms of spin-independent elastic scattering of WIMPs. The picture is summarized in Fig.~\ref{fig:direct}a.

\subsubsection{Recent Chronicle}
\begin{itemize}
\item {\it Since few years:}
{\bf DAMA} (and its successor DAMA/LIBRA) experiment (Bernabei et al., 2010) aims to investigate the annual modulation of WIMP signal given by the Earth's motion around the Sun with scintillating sodium iodide (NaI). They detected a signal with proper modulation features (i.e., cosine like with period $\sim1$ year and phase $\sim$ June 2nd), 
no modulation above 6 keV and in the 2-6 keV multiple-hits residual rate (WIMPs do not induce multiple-hits events), and high statistical significance ($\sim9\sigma$ C.L.). This makes hard to reconcile it with other sources of background and the DM interpretation is plausible. On the other hand, the "standard interpretation", i.e., spin-independent elastic scattering of WIMPs, is in tension with other experiments, unless the WIMP is very light ($m_{DM}\leq10$ GeV).~\footnote{More exotic scenarios may instead fit the DAMA excess without violating other constraints even for larger masses (see references in (Bernabei et al., 2010)).}

\item {\it December 2009:}
{\bf CDMS} (Ahmed et al., 2009) (Cryogenic Dark Matter Search) experiment measures phonons and ionization with Ge-Si detectors. They detected 2 events versus an expected background of 0.8 events, which means that the statistical significance (P=23\%, $1.5\sigma$) of the "excess" is too low to allow any conclusion. However, the interpretation in terms of WIMP scattering with $m_{DM}\sim10$ GeV is consistent with DAMA results (at "$2\sigma$").

\item {\it February 2010:}
{\bf COGENT} (Aalseth et al., 2010) detected a number of cosmogenic peaks, plus an exp-like signal not immediately identifiable with background. It is a Ge-detector with lower threshold than CDMS, but larger background contamination (detecting only the ionization signal cannot discriminate between nuclear and electron recoils), see Fig.~\ref{fig:direct}b. The region of interpretation in terms of WIMP scattering has some overlapping with DAMA and CDMS regions as shown in Fig.~\ref{fig:direct}a.

\item {\it May 2010:}
{\bf XENON100} (Aprile et al., 2010) collaboration with a setup made by liquid Xenon ($\sim100$ kg) aiming to reach ultra-low background, records recoil events through scintillation and ionization. They reported an analysis where no events has been detected excluding interpretation of above mentioned signals as being due to spin-independent, elastic, light mass WIMP scattering.

\item {\it Now (end of May 2010):}
The result of XENON100 has been immediately criticized in (Collar \& McKinsey, 2010a). The main issue relies on the fact that the relative scintillation efficiency $L_{eff}$ at low recoil energy, crucial for deriving constraints for low mass WIMPs, has been extrapolated due to the lack of experimental data for liquid Xenon below 4 keVr.
The XENON collaboration reaffirmed that their analysis relies on conservative assumptions (Xenon coll., 2010),  while (Collar \& McKinsey, 2010b) considered this reply inadequate and claimed that using a more "realistic" efficiency the constraints from current XENON100 analysis becomes less restrictive at low masses than bounds derived from XENON10 or CDMS data.
In the meantime, preliminary data from the CRESST experiment, including few events not immediately ascribable to background, have been leading to speculations (no official analysis has been performed by the team) showing that the parameter space for the interpretation of these data in terms of WIMPs can overlap with DAMA, CDMS, and COGENT (see Fig.~\ref{fig:direct}a (Schwetz, 2010)). 

\end{itemize}

As a word of caution, I highlight here that on top of experimental uncertainties, the extraction of constraints on the WIMP parameter space from events close to the threshold of a detector can be highly sensitive to the assumptions related to WIMP phenomenology, as e.g., the mean velocity, which maybe invoked to significantly shift the contours in Fig.~\ref{fig:direct}a.

Therefore, the conservative conclusion that can be drawn by this chronicle is that a common explanation for DAMA+CDMSII+COGENT (+CRESST) 
in terms of spin-independent, elastic scatterings of WIMPs with $m_{DM}\sim10$ GeV and $\sigma_s\sim 10^{-40}-10^{-41} {\rm cm}^2$  
is currently constrained but not completely ruled out.
This scenario does not match common naive expectations for WIMPs from, e.g., supersymmetry (although see (Bottino et al., 2010)), since requires a rather low mass and large scattering cross section. 

\subsection{Indirect Detection}
Indirect detection strategies involve signals associated to fluxes of particles originated from WIMP annihilations or decays in astrophysical structures. The search has been focussed on antimatter, photons, and neutrinos.

\subsubsection{Antimatter}
Positrons, anti-protons, and anti-deuterium, induced by WIMP annihilations/decays in the galactic halo, can be detected
as an exotic contribution in the local spectra of cosmic-rays (CRs).

Recent measurements of the positron fraction up to 100~GeV by the PAMELA experiment (Adriani et al., 2009) 
have triggered a lot of interest on the possibility that there may be a dominant  
contribution to the positron flux induced by WIMPs.
The data show a sharp raise at high-energy which is puzzling if the source is given by interactions of CRs with the interstellar medium (under "standard" propagation assumptions), while seems to indicate the presence of a nearby (within few kpc) source of positrons.
In order to ascribe the "excess" to WIMPs, the DM particle has to be heavy ($m_{DM}\geq100$ GeV or, probably, $\geq1$ TeV combining PAMELA data with recent FERMI $e^+-e^-$ data at low energy), leptophilic (i.e. with dominant branching ratio of annihilation into leptons), and with large annihilation rate.  
In other words, it requires a quite non-standard scenario. In this talk I do not focus on the related model building, but rather I try to answer to the question:
How can we distinguish between viable astrophysical (e.g., pulsars and supernova remnants) and dark matter interpretations of the excess?
Many tests have been suggested, and I restrict to the more robust ones, namely, involving the same channel (electrons/positrons) and the same region (the local part of the Milky Way) as for the positron excess.
Anisotropy in the $e^+-e^-$ spectrum and bumpiness of the $e^+$ spectrum have been proposed, and can be expected to be sizable in case of dominant contribution from pulsars (Hooper, Blasi \& Serpico, 2009) or DM substructures (Regis \& Ullio, 2009b). On the other hand, they are quite challenging measurements and a null result wouldn't shed light on the issue.
The radiative diffuse emission at mid-latitudes is a probe of the CR population in the nearby region. In particular, the gamma-ray inverse Compton emission is originated by high-energy electrons and positrons. 
A different spatial distribution is expected if they are induced by DM annihilations or decay in the halo (an extended nearly spherical profile)
rather than by astrophysical sources (confined within the stellar disc), so this two emissions can be disentangled (Regis \& Ullio, 2009a). Another advantage of this test relies on the fact that it is based on {\it local} quantities, such as DM distribution and propagation parameters, whose determination is much more robust than in other regions, e.g., Galactic center. 
The possibility of fully performing this test will come with forthcoming data of the Fermi Telescope at energies above 100 GeV and $|b|\geq 10^\circ$ as shown in Fig~\ref{fig:indirect}a ((Regis \& Ullio, 2009a) and (Cumberbatch et al., 2010)).
A discovery of a $\gamma$-ray term, with spectral and angular features as for the DM source would be a striking signature.
Conversely, an observational result in agreement with the prediction from standard CR components only would imply very strong constraints on  
the DM interpretation of the PAMELA excess.

\subsubsection{Photons}
\begin{figure}  

 \begin{minipage}[htb]{5cm}
   \centering
   \includegraphics[width=5cm]{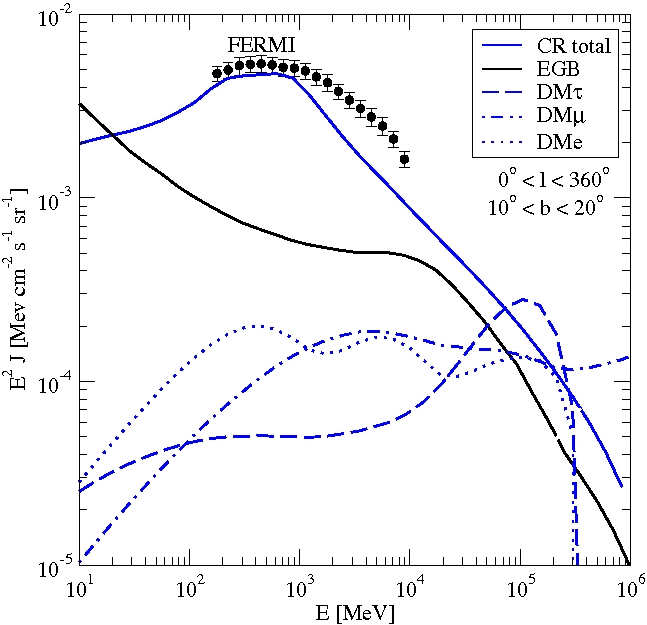}
 \end{minipage}
 \ \hspace{1cm} \
\begin{minipage}[htb]{6cm}
\includegraphics[width=6cm]{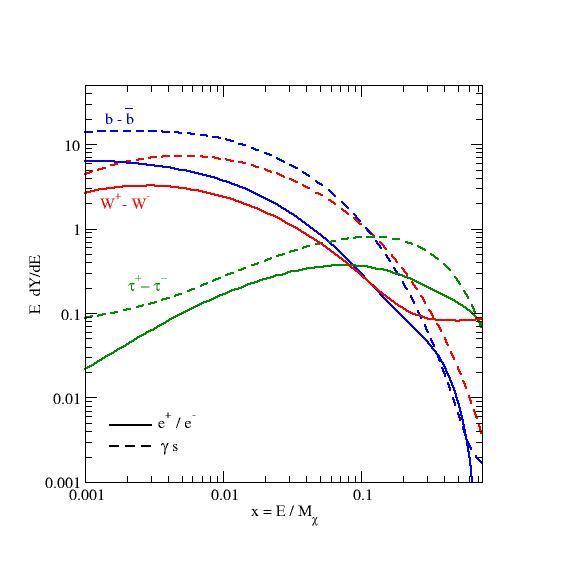}
 \end{minipage}
\caption{{\it Left:} Contribution to diffuse emission at intermediate latitudes for three different WIMP models (dashed, dotted, and dashed-dotted lines) fitting the PAMELA excess. Experimental data points and benchmark galactic and extra-galactic backgrounds are also shown. From  (Regis \& Ullio, 2009a).
{\it Right:} Photon and electron annihilation yield for three different final states of WIMP annihilation. From  (Regis \& Ullio, 2008). }
\label{fig:indirect}
\end{figure}

Prompt emissions of gamma-ray photons in the DM halo include continuum emission proceeding through the production and decay of neutral pions and final state radiation and monoenergetic spectral signatures related to direct production at loop level.
Electrons and positrons can be directly or indirectly produced by WIMP annihilations. 
They act as sources for radiative processes generating a multi-wavelength spectrum, mainly involving synchrotron emission at radio/infrared frequencies and inverse Compton (IC) scattering on CMB and starlight in the X- and gamma-ray bands .
In Fig.~\ref{fig:indirect}, we show the WIMP final yield in photons and electrons for three completely different channels of annihilation. In all cases, the two yields are comparable which means that, in general, a comparable luminosity at different wavelengths is expected and therefore that multi-wavelength searches can be a very powerful technique in WIMP searches (Regis \& Ullio, 2008).

The search for prompt and radiative emissions of gamma-ray photons from WIMP annihilations and decays constitutes one of the prime goals of the FERMI collaboration, e.g., (Murgia, 2010).
The objects of investigation include: Galactic center, Milky Way halo, extra-galactic regions, satellites (dwarf spheroidal and DM substructures), and clusters of galaxies, with focuses on both spectral line and continuum signatures.
The current (after 1.5 year of mission) main conclusion from the team is that there has been no discovery so far, but, however, constraints on the nature of DM have been significantly improved. For the near future, prospects are still very promising since the ongoing analysis for a better understanding of the detector
response and of astrophysical backgrounds will significantly improve the potential for DM discovery.

Other authors pointed out that possible hints of DM maybe actually already contained in the FERMI data.
In particular, a diffuse inverse Compton signal in the inner Galaxy with a similar spatial morphology of the so called WMAP "haze" (Hooper, Finkbeiner \& Dobler, 2007) have been claimed in (Dobler et al., 2009). The morphology and spectrum are consistent with it being the IC counterpart of the synchrotron microwave signal.
However, the absence of an analogous excess in the polarization data at microwave, systematics in gamma-ray background templates (Linden \& Profumo, 2010), and, most importantly, great uncertainties in modeling both DM and CR in the inner Galaxy make the interpretation of the emission and the possible link with DM microscopic properties quite challenging.

\subsubsection{Neutrinos}

The most favourable targets for detecting neutrinos induced by DM annihilations/decays are Galactic center, Sun (most promising) and Earth. 
The neutrino flux depends strongly on the WIMP capture, and in turn on elastic cross section with light nuclei, rather than only on the annihilation cross section as in the previous indirect detection methods. Therefore this strategy provides constraints complementary to direct and indirect searches.
Prospects for detection in a km-size neutrino telescope, such as IceCube, are intriguing (Bergstrom, Edsjo \& Gondolo, 1998).

\section{Summary and Conclusions}
Recent "excesses" in direct detection experiments and in the PAMELA detectors share few properties: 
a) WIMP interpretations require rather non-standard scenarios and 2) are constrained by other messengers/experiments; 3) other viable explanations exist. 
Their main difference is instead related to the WIMP mass required to fit the excess, which is light in the direct detection case (when combining all the experimental results mentioned in Sec.~3.2), while heavy in the PAMELA case. This is a challenging but also intriguing aspect for a unified theoretical picture.

In conclusion, I didn't answer to the initial question about when and through which method the DM will be unambiguously discovered.
On the other hand, I can safely state that we have a very well-motivated class of DM models (i.e., WIMPs) which different techniques are starting to test, as also the recent history has been showing, and which will be definitively probed in the forthcoming years/decade.

\section{Acknowledgements}
This work is supported by the NRF and CHPC (South Africa).





\begin{thebibliography}{99}

\bibitem{Aalseth:2010vx}
  C.~E.~Aalseth {\it et al.}  [CoGeNT collaboration],
  arXiv:1002.4703 

\bibitem{Adriani:2008zr}
  O.~Adriani {\it et al.}  [PAMELA Collaboration],
  Nature {\bf 458} (2009) 607.

\bibitem{Ahmed:2009zw}
  Z.~Ahmed {\it et al.}  [The CDMS-II Collaboration],
  arXiv:0912.3592 

\bibitem{Aprile:2010um}
  E.~Aprile {\it et al.}  [XENON100 Collaboration],
  arXiv:1005.0380 

\bibitem{Bergstrom:1998xh}
  L.~Bergstrom, J.~Edsjo and P.~Gondolo,
  Phys.\ Rev.\  D {\bf 58} (1998) 103519


\bibitem{Bernabei:2010ke}
  R.~Bernabei {\it et al.},
these proceedings,
  arXiv:1007.0595. 

\bibitem{Bottino:2009km}
  A.~Bottino et al.,
  Phys.\ Rev.\  D {\bf 81} (2010) 107302

\bibitem{Bosma:1981}
 Bosma,~A.
 1981 AJ 86, 1791.

\bibitem{Clowe:2006eq}
  D.~Clowe et al.,
  Astrophys.\ J.\  {\bf 648} (2006) L109

\bibitem{Collar:2010gg}
  J.~I.~Collar and D.~N.~McKinsey,
  arXiv:1005.0838 

\bibitem{Collar:2010gd}
  J.~I.~Collar and D.~N.~McKinsey,
  arXiv:1005.3723 

\bibitem{Cumberbatch:2010ii}
  D.~T.~Cumberbatch et al.,
  arXiv:1003.2808 

\bibitem{Dobler:2009xz}
  G.~Dobler et al.,
  Astrophys.\ J.\  {\bf 717} (2010) 825

\bibitem{Feng:2010gw}
  J.~L.~Feng,
  arXiv:1003.0904. 


\bibitem{Hooper:2007kb}
  D.~Hooper, D.~P.~Finkbeiner and G.~Dobler,
  Phys.\ Rev.\  D {\bf 76} (2007) 083012

\bibitem{Hooper:2008kg}
  D.~Hooper, P.~Blasi and P.~D.~Serpico,
  JCAP {\bf 0901} (2009) 025

\bibitem{Jungman:1995df}
  G.~Jungman et al.,
  Phys.\ Rept.\  {\bf 267} (1996) 195

\bibitem{Kolb:1990vq}
  E.~W.~Kolb and M.~S.~Turner,
  Front.\ Phys.\  {\bf 69} (1990) 1.

\bibitem{Komatsu:2010fb}
  E.~Komatsu {\it et al.},
  arXiv:1001.4538 

\bibitem{Linden:2010ea}
  T.~Linden and S.~Profumo,
  Astrophys.\ J.\  {\bf 714} (2010) L228

\bibitem{fermi}
S.~Murgia talk at GGI conference,
http://www.ggi.fi.infn.it/

\bibitem{Panico:2008bx}
  G.~Panico et al.,
  Phys.\ Rev.\  D {\bf 77} (2008) 115012

\bibitem{Regis:2006hc}
  M.~Regis, M.~Serone and P.~Ullio,
  JHEP {\bf 0703} (2007) 084

\bibitem{Regis:2008ij}
  M.~Regis and P.~Ullio,
  Phys.\ Rev.\  D {\bf 78} (2008) 043505

\bibitem{Regis:2009md}
  M.~Regis and P.~Ullio,
  Phys.\ Rev.\  D {\bf 80} (2009) 043525

\bibitem{Regis:2009qt}
  M.~Regis and P.~Ullio,
  arXiv:0907.5093 



\bibitem{straessner}
Arno Straessner,
these proceedings.


\bibitem{schwetz}
T.~Schwetz talk at GGI conference,
http://www.ggi.fi.infn.it/

\bibitem{Tisserand:2006zx}
  P.~Tisserand {\it et al.}  [EROS-2 Collaboration],
  Astron.\ Astrophys.\  {\bf 469} (2007) 387

\bibitem{Collaboration:2010er}
  The Xenon Collaboration,
  arXiv:1005.2615 

\bibitem{Zwicky:1933}
F.~Zwicky, 
Helv. Phys. Acta {\bf 6} (1933) 110.

\end{thebibliography}
\end{document}